%% file: main.tex
\documentclass[a4paper]{article}
\usepackage{INTERSPEECH2020}

\usepackage{todonotes}
\usepackage{multirow}
\usepackage{subcaption}
\usepackage{tipa}

\title{THE ICASSP SP CADENZA CHALLENGE: MUSIC DEMIXING/REMIXING FOR HEARING AIDS}
\name{
Gerardo Roa Dabike$^1$, 
Michael A. Akeroyd$^2$,
Scott Bannister$^3$, 
Jon Barker$^4$,\\
Trevor J. Cox$^1$,
Bruno Fazenda$^1$,
Jennifer Firth$^2$,
Simone Graetzer$^1$, 
Alinka Greasley$^3$,\\
Rebecca R. Vos$^1$,
William M. Whitmer$^2$
}
\address{
University of Salford$^1$,
University of Nottingham$^2$, 
University of Leeds$^3$, 
University of Sheffield$^4$}
\email{cadenzachallengecontact@gmail.com}

\begin{document}
\maketitle

\begin{table}[t]

\copyright\ 2024 IEEE. Personal use of this material is permitted. Permission from IEEE must be obtained for all other uses, in any current or future media, including reprinting/republishing this material for advertising or promotional purposes, creating new collective works, for resale or redistribution to servers or lists, or reuse of any copyrighted component of this work in other works.
\end{table}

\input{0-abstract}
\input{1-introduction}
\input{2-description}

\input{3-Submissions}
\input{5-conclusions}
\input{6-acknowledge}

\bibliographystyle{IEEEtran}

\bibliography{IEEEabrv, main}

\end{document}

%% file: 0-abstract.tex
\begin{abstract}
    This paper reports on the design and results of the 2024 ICASSP SP Cadenza Challenge: Music Demixing/Remixing for Hearing Aids. The Cadenza project is working to enhance the audio quality of music for those with a hearing loss. The scenario for the challenge was listening to stereo reproduction over loudspeakers via hearing aids. The task was to: decompose pop/rock music into vocal, drums, bass and other (VDBO); rebalance the different tracks with specified gains and then remixing back to stereo. End-to-end approaches were also accepted. 17 systems were submitted by 11 teams. Causal systems performed poorer than non-causal approaches. 9 systems beat the baseline. A common approach was to fine-tuning pretrained demixing models. The best approach used an ensemble of models.
\end{abstract}

%% file: 1-introduction.tex
\section{Introduction}

430 million people worldwide experience disabling hearing loss, with unaddressed hearing loss having a global cost of US \$980 billion per annum\cite{WHO}. Hearing loss has a number of effects, including making it harder to pick out sounds from a mixture, such as the melody line from a band. This can make music less enjoyable and risks people disengaging from listening and creating music.

Hearing aids are the main treatment for hearing loss. The default settings on hearing aids are optimized for speech, however. 68\% of users report difficulties when listening to music through hearing aids \cite{Greasley2020}; the effectiveness of music programmes on hearing aids varies.

The Cadenza Project is addressing this need for better music processing through  machine learning challenges. The first Cadenza Challenge (CAD1) \cite{CAD1} started in 2023. The subsequent 2024 ICASSP Cadenza Challenge\footnote{https://cadenzachallenge.org/} built upon the CAD1 headphone task.

%% file: 2-description.tex
\section{ICASSP challenge Description}


A person is listening to music over stereo loudspeakers. The signals to be processed are from the hearing aid microphones at each ear. Entrants were challenged to rebalance the vocal, drums, bass and other (VDBO) components by specified gains. Such a system would then allow music to be personalised, for example, amplifying the vocal component to increase lyric intelligibility.

\begin{figure}[!t]
    \centering
    \includegraphics[width=0.4\linewidth]{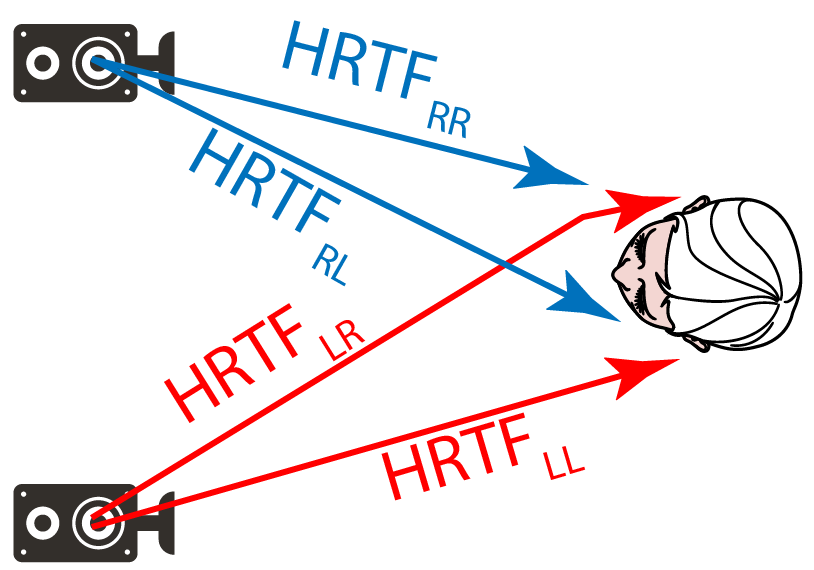}
    \caption{Left and right signals mixed using anechoic Head Related Transfer Functions (HRTFs)}
    \label{fig:crosstalk}
\end{figure}

The microphone signals were a mixture of both the right and left loudspeaker signals -- see Figure \ref{fig:crosstalk}. Head-Related Transfer Functions (HRTFs) modelled the sound propagation from the loudspeakers to the hearing aid microphones. These were from OlHeaD-HRTF \cite{Denk2018}. The mixing of the left-right loudspeaker signals was strongest at low frequencies where wavelengths were large compared to head size. Thus, the left-right balance of VDBO components differed in the hearing aid signals compared to the original loudspeaker feeds. Entrants who used a demix/remix approach, therefore faced additional challenge compared to CAD1 and previous demixing challenges.

\begin{figure}[!t]
    \centering
    \includegraphics[width=\linewidth]{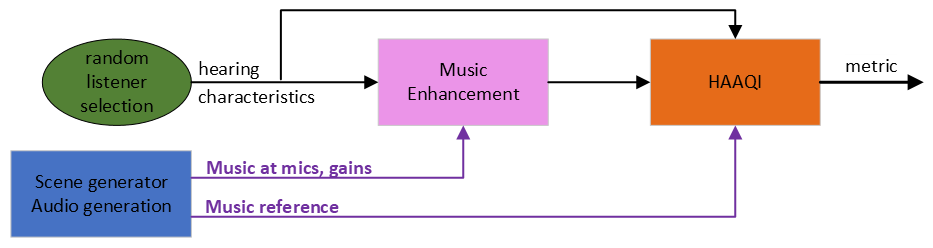}
    \caption{Schematic of the challenge baseline}
    \label{fig:Baseline}
\end{figure}

Figure \ref{fig:Baseline} shows a schematic of the challenge baseline. A scene generator (blue box) randomly generated scene characteristics, which included selecting the music track, choosing HRTFs characterized by the loudspeaker locations, and selecting one of the 16 subjects from the OlHeaD-HRTF dataset. Additionally, it determined the gains to be applied to each VDBO stem in the downmix. The listener audiograms were provided as metadata (green oval).

The music enhancement stage (pink box) was where an entrants' system was placed. This took the music captured by the hearing aid microphones as inputs. The output processed signals were the rebalanced stereo music.

The remixed stereo was evaluated using the Hearing-Aid Audio Quality Index (HAAQI) \cite{kates2015hearing}. The reference for this intrusive metric was the rebalanced stereo using the ground truth VDBO with the appropriate HRTFs, gains and audiograms applied. 

HAAQI allows for the raised hearing threshold of listeners in its calculation. The thresholds came from bilateral pure-tone audiograms at [250, 500, 1000, 2000, 3000, 4000, 6000, 8000] Hz. A broader bandwidth would have been preferable for music, but was not possible due to available databases and the gain rules used in HAAQI. There were 83, 51 and 53 independent pairs of measured audiograms for the training, validation and evaluation sets.

The music for training and evaluation used the standard splits from MUSDB18-HQ \cite{musdb18-hq}, corresponding to 100 and 50 stereo tracks, respectively. An independent validation set was constructed by randomly selecting 50 tracks from  MoisesDB \cite{moisesdb}, maintaining the same genre distribution as the evaluation split of MUSDB18-HQ. This was done because many pretrained models that use MUSDB18-HQ, incorporated the validation split as part of their training.




The two baseline systems (T01\&02) were out-of-the-box pretrained audio source separation networks (with no retraining to allow for the loudspeaker scenario). T01 used the Hybrid Demucs model \cite{defossez2021hybrid}, which employs a U-Net architecture to combine both time-domain and spectrogram-based audio source separation. T02 used the Open-Unmix model \cite{stoter19}, which just uses spectrograms.

%% file: 3-Submissions.tex
\section{Submissions and Results}

There were 17 systems from 11 teams - see Table \ref{tab:results}. Systems that employed data augmentation or supplementation were scored before and after applying these techniques. Nearly all differences between the system scores in Table \ref{tab:results} are statistically significant, but some have very small effect sizes. 9 systems beat the best baseline (T01). Systems T22 and T47 scored higher using an ensemble of fine-tuned pretrained systems.




%% file: 5-conclusions.tex
\section{Discussion and Conclusions}

Existing audio source separation networks like HDemucs from the Baseline T01, needed to be adapted to work with the microphone signals at the ear (as was done by T09, T09B, T11, T46). Music source separation was more difficult for this scenario, however, due to the frequency-dependent mixing of the VDBO components across the left-right channels. 

The entrants' submissions comprised different techniques, ranging from ensembles of two or more audio source separation networks (T47), to the use of traditional machine learning techniques (T16). When applied with care, techniques like data augmentation and data supplementation were found to aid model generalisation (T03, T11, T31, and T42), but the gains in HAAQI were modest. Another insight from T18, was that extracting instrument-based sub- and full-band information in conjunction with beamforming can help.  

In the long run, a system that is deployable on a hearing-aid needs to have a small model size, be causal and be personalised to a listeners' hearing acuity.

System T16 employed traditional machine learning technologies that required lower resources. Listener's ear embedding was explored by T09, letting the model learn to apply the amplification. T09, T09B and T16 used causal approaches. These all performed worse than all the non-casual approaches, however.

Future audio source separation challenges therefore need to encourage causal and low-latency approaches. This would enable those with hearing loss to benefit from the latest audio machine learning.

Future Cadenza challenges will also address other issues that listeners using hearing aids have with music, such as coping with large dynamic ranges without introducing distortion.

\begin{table}[!t]
    \centering
    \caption{System results for the evaluation set. *:  causal system. A: data augmentation. S: supplementary data. B: second submission.}
\scalebox{0.92}{
    \begin{tabular}[t]{l|c}
    \toprule
         \textbf{Entry}	& \textbf{HAAQI} \\
    \midrule
         T47 \cite{T022} & \textbf{0.632} \\
         T22 & \textbf{0.631} \\
         T03S \cite{T003} & \textbf{0.593} \\
         T03 \cite{T003} & \textbf{0.592} \\
         T11A \cite{T011} & \textbf{0.586} \\
         T18 \cite{T018} & \textbf{0.585} \\
         T11 \cite{T011} & \textbf{0.580} \\
    \bottomrule
    \end{tabular}
    \hspace{0.3em}
    \begin{tabular}[t]{l|c}
    \toprule
         \textbf{Entry}	& \textbf{HAAQI} \\
    \midrule
         T12 & \textbf{0.573} \\
         T46 \cite{T046} & \textbf{0.570} \\
         T01 & 0.570 \\
         T25 & 0.561 \\
         T31A & 0.543 \\
         T42 & 0.543 \\
         T42A & 0.534 \\
    \bottomrule
    \end{tabular}

    \hspace{0.3em}
    \begin{tabular}[t]{l|c}
    \toprule
         \textbf{Entry}	& \textbf{HAAQI} \\
    \midrule
         T31 & 0.530 \\    
         T02 & 0.511 \\
         T09B* & 0.479 \\         
         T09* & 0.478 \\
         T16* & 0.144 \\
    \bottomrule
    \end{tabular}
}
     
    \label{tab:results}
\end{table}





%% file: 6-acknowledge.tex
\section{Acknowledgements}
     Cadenza is funded by the Engineering and Physical Sciences Research Council (EPSRC) [EP/W019434/1]. We thank our partners: BBC, Google, Logitech, RNID, Sonova, Universit\"{a}t Oldenburg.